\documentclass[12pt,epsf]{article}
\usepackage{graphicx}
\usepackage{amsmath,amssymb}
\begin{document}
%%%%%%%%%%%%%%%%%%%%%%%%%%%%%%%%%%%%%%%%%%%

\def\a{\alpha}
\def\b{\beta}
\def\c{\varepsilon}
\def\d{\delta}
\def\e{\epsilon}
\def\f{\phi}
\def\g{\gamma}
\def\h{\theta}
\def\k{\kappa}
\def\l{\lambda}
\def\m{\mu}
\def\n{\nu}
\def\p{\psi}
\def\q{\partial}
\def\r{\rho}
\def\s{\sigma}
\def\t{\tau}
\def\u{\upsilon}
\def\v{\varphi}
\def\w{\omega}
\def\x{\xi}
\def\y{\eta}
\def\z{\zeta}
\def\D{\Delta}
\def\G{\Gamma}
\def\H{\Theta}
\def\L{\Lambda}
\def\F{\Phi}
\def\P{\Psi}
\def\S{\Sigma}

\def\o{\over}
\def\beq{\begin{eqnarray}}
\def\eeq{\end{eqnarray}}
\newcommand{\gsim}{ \mathop{}_{\textstyle \sim}^{\textstyle >} }
\newcommand{\lsim}{ \mathop{}_{\textstyle \sim}^{\textstyle <} }
\newcommand{\vev}[1]{ \left\langle {#1} \right\rangle }
\newcommand{\bra}[1]{ \langle {#1} | }
\newcommand{\ket}[1]{ | {#1} \rangle }
\newcommand{\EV}{ {\rm eV} }
\newcommand{\KEV}{ {\rm keV} }
\newcommand{\MEV}{ {\rm MeV} }
\newcommand{\GEV}{ {\rm GeV} }
\newcommand{\TEV}{ {\rm TeV} }
\def\diag{\mathop{\rm diag}\nolimits}
\def\Spin{\mathop{\rm Spin}}
\def\SO{\mathop{\rm SO}}
\def\O{\mathop{\rm O}}
\def\SU{\mathop{\rm SU}}
\def\U{\mathop{\rm U}}
\def\Sp{\mathop{\rm Sp}}
\def\SL{\mathop{\rm SL}}
\def\tr{\mathop{\rm tr}}

\def\IJMP{Int.~J.~Mod.~Phys. }
\def\MPL{Mod.~Phys.~Lett. }
\def\NP{Nucl.~Phys. }
\def\PL{Phys.~Lett. }
\def\PR{Phys.~Rev. }
\def\PRL{Phys.~Rev.~Lett. }
\def\PTP{Prog.~Theor.~Phys. }
\def\ZP{Z.~Phys. }

%%%%%%%%%%%%%%%%%%%%%%%%%%%%%%%%%%%%%%%%%%%%%%%%%%%%%%%%%%%%%%%%%%%%

\baselineskip 0.7cm

\begin{titlepage}

\begin{flushright}
IPMU 15-0010\\
\end{flushright}

\vskip 1.35cm
\begin{center}
{\large \bf Higgs mass 125 GeV and $g-2$ of the muon in Gaugino Mediation Model
}
\vskip 1.2cm
  Keisuke Harigaya$^{a}$,  Tsutomu T. Yanagida$^{a}$ and Norimi Yokozaki$^{b}$
\vskip 0.4cm

{\it       
$^a$ Kavli IPMU (WPI), TODIAS, the University of Tokyo, Kashiwa, Chiba 277-8583, Japan \\
$^b$ INFN, Sezione di Roma, Piazzale A. Moro 2, I-00185 Roma, Italy
}

\vskip 1.5cm

\abstract{
Gaugino mediation is very attractive since it is free from the serious flavor problem in the supersymmetric standard model.
We show that the observed Higgs boson mass at around 125 GeV and the anomaly of the muon $g-2$ can be easily explained in gaugino mediation models.
It should be noted that no dangerous CP violating phases are generated in our framework.
Furthermore, there are large parameter regions which can be tested not only at the planned International Linear Collider but also at the coming 13-14 TeV Large Hadron Collider.  
}

\end{center}
\end{titlepage}

\setcounter{page}{2}

\section{Introduction}

The minimal supersymmeric standard model (MSSM) is an attractive candidate for the physics beyond the standard model (SM). The gauge coupling unification is beautifully realized with new particles predicted by supersymmetry (SUSY). The longstanding anomaly of the muon $g-2$~\cite{exp_gm2} (see  Refs.~\cite{gm2_sm1, gm2_sm2} for SM predictions) is resolved if sleptons, Bino and Wino exist at around the weak scale~\cite{susy_gm2,susy_gm2_2}. Thus, low-energy SUSY, containing light SUSY particles, is especially attractive.  However, there is a serious obstacle in low-energy SUSY models: the SUSY flavor problem. Without the suppression of  flavor violating soft masses, sleptons need to be much heavier than the weak scale, otherwise non-observed flavor violating decays, e.g., $\mu \to e \gamma$ are generated with large branching fractions~\cite{susy_meg}. Obviously, these heavy sleptons are inconsistent with the explanation of the muon $g-2$ anomaly. Therefore, we need a SUSY breaking model which largely suppresses flavor violating soft masses.

Gaugino mediation \cite{Inoue:1991rk,gaugino_med} provides a convincing solution to the SUSY flavor problem, where soft masses of squarks and sleptons vanish at a high energy scale. At low energy scales, these soft masses are generated radiatively by gaugino loops and hence they are flavor-blind. As a result, gaugino mediation is free from the serious SUSY flavor problem.\footnote{
Also, gaugino mediation provides an attractive solution to the fine-tuning problem of the electroweak symmetry breaking scale:~Focus point gaugino mediation~\cite{focus}. In focus point gaugino mediation, the electroweak symmetry scale is naturally explained by the SUSY particle mass scale even for few TeV gaugino masses.
} 

In this paper, we show that the observed Higgs boson mass $m_h\simeq 125$ GeV~\cite{atlas_higgs, cms_higgs} as well as the anomaly of the muon $g-2$ can be easily explained in our framework of gaugino mediation, although there is a  tension between the muon $g-2$ and the observed Higgs boson mass in general.
Here, in our framework,  we assume that two Higgs doublets couple to a SUSY breaking field, allowing enhancements of a SUSY contribution to the muon $g-2$ with a light Higgsino and radiative corrections to the Higgs boson mass through a sizable left-right mixing of the stops. 
%In fact,  %the muon $g-2$ anomaly suggests the SUSY particles exist at around the weak scale, while 
%$m_h\simeq 125$ GeV~\cite{higgs_exp_mass} requires large radiative corrections from heavy stops as $\mathcal{O}(1-10)$\,TeV~\cite{}, which makes it difficult to enhance the SUSY contributions to the muon $g-2$.
We also discuss the possibility of a large gravitino mass in comparison with MSSM soft masses,
which relaxes cosmological problems.

\section{Gaugino mediation model}

\label{sec:gaugino med}

We consider the Kahler potential such that squarks and sleptons masses vanish at the tree level.
One example is so-called the sequestering~\cite{Inoue:1991rk},
which may be realized in the brane world~\cite{Randall:1998uk}.
Another example is the SUSY non-linear sigma model~\cite{Bardeen:1981df,Buchmuller:1982xn,Buchmuller:1982tf}
where squarks and sleptons are assumed be pseudo-Nambu-Goldstone bosons.
Due to the Nambu-Goldstone-nature of squarks and sleptons, soft masses of them vanish at the tree-level.
As long as low energy phenomenology is concerned, 
the following discussion based on the sequestered Kahler potential
is essentially the same as on the Kahler potential in SUSY non-linear sigma models.%
 \footnote{{ There would be differences when one discusses quantum corrections to soft masses at a high energy scale.
 (See the discussion in Appendix~\ref{sec:soft mass}.)
 }}
Thus, we discuss the sequestered Kahler potential as an example.
We comment on SUSY non-linear sigma models, if necessary.

The Kahler potential and the super potential are given by
\begin{eqnarray}
K  = -3 M_P^2 \log\Bigl[1 - \frac{f(Z+Z^\dag) + Q_i^\dag Q_i +\Delta K}{3 M_P^2} \Bigr] , \ \ W = \mathcal{C} + \tilde{\mu} H_u H_d + W_{\rm Yukawas},
\label{eq:sequestering}
\end{eqnarray}
where $M_P$ is the reduced Planck scale, $Z$ is a chiral superfield, $f$ is a real function, $Q_i$ are MSSM chiral superfields, $\mathcal{C}$ is a constant, and $\tilde{\mu}$ is the Dirac mass term of Higgsinos. $\Delta K$ is explained later.
For an appropriate choice of the function $f$, SUSY is broken by the $F$ term of $Z$ and the cosmological constant vanishes at the vacuum, without introducing a linear term of $Z$ in the super potential~\cite{Izawa:2010ym}.
(See Ref.~\cite{Cremmer:1983bf} for the case of $f(Z+Z^\dag$) linear in $Z + Z^\dag$.)
A concrete example of $f$ is given in Ref.~\cite{Iwamoto:2014ywa}.
In the following, we assume that SUSY is dominantly broken by the $F$ term of $Z$ and call $Z$ as the SUSY breaking field. We refer to this type of SUSY breaking as the gravitational SUSY breaking, following Ref.~\cite{Izawa:2010ym}.

In addition to the sequestering,
we have assumed the shift symmetry, $Z \rightarrow Z + i r$ with a real constant $r$,
so that no CP violating phases arise~\cite{Iwamoto:2014ywa}.%
\footnote{This shift symmetry automatically arises in SUSY non-linear sigma models~\cite{Komargodski:2010rb,Kugo:2010fs}.}
Actually,
%by a Peccei-Quinn rotation
by phase rotations of MSSM fields
and an $U(1)_R$ rotation,
we can take $\tilde{\mu}$ and $\mathcal{C}$ real.
Due to the shift symmetry and the reality of the Kahler potential, all couplings in the Kahler potential (and gauge kinetic functions, as we will see,) are real.
Thus, physical CP phases vanish in our setup.

We assume that the up- and down-type Higgs couple to the SUSY breaking field through the Kahler potential;%
\footnote{If the Kahler potential contains a term $K \supset c H_uH_d$, the constant $c$ is in general complex after taking $\tilde{\mu}$ real.
If $c=O(1)$, this term leads to large CP violations.
However, $c$ is suppressed unless the combination $H_u H_d$ has vanishing charges under any symmetries.}
\begin{eqnarray}
\Delta K &=& [ c_{u} \frac{(Z+Z^\dag)}{M_P} + d_{u} \frac{(Z+Z^\dag)^2}{M_P^2} + \cdots] H_u^\dag H_u  \nonumber \\
&&+  [ c_{d} \frac{(Z+Z^\dag)}{M_P} + d_{d} \frac{(Z+Z^\dag)^2}{M_P^2} + \cdots] H_d^\dag H_d,
\label{eq:deltaK}
\end{eqnarray}
where $c_u$, $c_d$, $d_u$, and $d_d$ are real constants and ellipces denote terms higher order in $Z + Z^\dag$.
In the sequestering scenario based on the brane world, these couplings can be understood by assuming that Higgs doublets live in bulk.
%In SUSY non-linear sigma models, they can be understood if higgses are not pseudo-Nambu-Goldstone bosons.
%Note that these couplings are not relevant to flavor violating soft masses.
%where $r_u$ and $r_d$ are introduced such that $H_u$ and $H_d$ are canonically normalized: $r_{u,d}=(1 + c_{u,d} x  + d_{u,d} x^2)/(1-f/(3M_P^2))$. ($x=\left<Z + Z^*\right>/M_P$).

Hereafter, we shift $Z$ by a constant so that the vacuum expectation value (VEV) of $Z + Z^\dag$ vanishes, and regard $Z$ in $\Delta K$ as the shifted field.
Then chiral fields with canonical kinetic terms are given by
\begin{eqnarray}
Q_i^c = \left(1-\vev{f}/3M_P^2\right)^{-1/2}Q_i,
\end{eqnarray}
where $\vev{\cdots}$ denotes the VEV of $\cdots$.
The Higgsino mass is then given by
\begin{eqnarray}
\mu = e^{\vev{K}/2M_P^2}\left(1-\vev{f}/3M_P^2\right) \tilde{\mu}.
\end{eqnarray}
For simplicity, we omit $\vev{}$ in the following.

The scalar soft mass squared for $H_u$ and $H_d$ are given by (see Appendix~\ref{sec:soft mass})
\begin{eqnarray}
%m_{H_u}^2 &=& e^{K/M_P^2} \Bigl[c_u'^2(1-f/3)^{-1} - d_u'  \Bigr] k^2 \frac{9|\mathcal{C}|^2}{M_P^4}, \nonumber \\
%m_{H_d}^2 &=& e^{K/M_P^2} \Bigl[c_d'^2(1-f/3)^{-1} - d_d'  \Bigr] k^2 \frac{9|\mathcal{C}|^2}{M_P^4}, \label{eq:soft}
m_{H_u}^2 &=&
9 k^2 (1-f/3M_P^2)^2  \Bigl[-2 d_u + c_u^2   \Bigr] \times m_{3/2}^2
  , \nonumber \\
m_{H_d}^2 &=& 
9 k^2 (1-f/3M_P^2)^2  \Bigl[-2 d_d +  c_d^2  \Bigr] \times m_{3/2}^2,
\label{eq:soft}
\end{eqnarray}
where $m_{3/2}$ is the gravitino mass.
Higgs trilinear couplings and the $B_\mu$-term are
\begin{eqnarray}
 A_u &=& -3k (1-f/3M_P^2) c_u \times m_{3/2} , \nonumber \\
 A_d&=& -3k (1-f/3M_P^2) c_d \times m_{3/2}, \nonumber \\
 B_{\mu} &=&  (A_u + A_d )  \mu,
 \label{eq:ab-terms}
\end{eqnarray}
where the constant $k$ is given by 
\begin{eqnarray}
k = \left(\frac{\q f}{\q x} \right)^{-1}.
\end{eqnarray}
In SUSY non-linear sigma models, those soft masses in general exist if Higgs doublets are not Nambu-Goldstone bosons.

Note that the $B_\mu$-term is proportional to the sum of trilinear couplings, and vanishes when MSSM higgses are also sequestered from the SUSY breaking field (i.e.~$c_u= d_u = c_d = d_d =0$).
This is not the case for generic sequestering scenarios, because of the VEV of the scalar auxiliary component of the supergravity multiplet.
(In the conformal formulation, it is the VEV of the $F$ term of the compensator.)
In the gravitational SUSY breaking, the VEV vanishes at the vacuum~\cite{Izawa:2010ym} and hence the $B_\mu$-term vanishes in the sequestered limit.

Next, we consider gaugino masses. As a model of the grand unified theory (GUT), we consider the $SU(5) \times SU(3)_H \times U(1)_H$ product group unification (PGU) model~\cite{Yanagida:1994vq,Hotta:1995cd}. In the PGU model, the doublet-triplet splitting problem, which is a serious problem in the minimal $SU(5)$ GUT, is solved. The gauge coupling unification is approximately maintained if the gauge coupling of $SU(3)_H \times U(1)_H$ is sufficiently stronger than that of $SU(5)$.

Relevant part of the Lagrangian is given by
\begin{eqnarray}
\mathcal{L} &\supset& \int d^2 \theta \Bigr[ \Bigr( \frac{1}{4g_5^2} - \frac{k_5 Z}{M_P} \Bigl)W_5 W_5 + \Bigr( \frac{1}{4g_{3H}^2} - \frac{k_{3H} Z}{M_P} \Bigl)W_{3H} W_{3H} \nonumber \\
&&+  \Bigr( \frac{1}{4g_{1H}^2} - \frac{k_{1H} Z}{M_P} \Bigl)W_{1H} W_{1H} \Bigl] + {\rm h.c.} \,.
\label{eq:gauge kin}
\end{eqnarray}
Here, $g_5$, $g_{3H}$ and $g_{1H}$ are the gauge coupling constants of $SU(5)$, $SU(3)_H$ and $U(1)_H$ gauge interactions, respectively. $W_5$, $W_{3H}$ and $W_{1H}$ are superfield field strength of each gauge multiplet.
$k_5$, $k_{3H}$ and $k_{1H}$ are constants, which are real in order to preserve the shift symmetry of $Z$.%
\footnote{The shift symmetry, $Z \rightarrow Z + i r $,  {has a quantum gauge anomaly and hence the shift symmetry is maintained in a perturbative limit.}}

After $SU(5) \times SU(3)_H\times U(1)_H$ is broken to $SU(3)_C \times SU(2)_L \times U(1)_Y$, non-universal gaugino masses are generated at the GUT scale~\cite{ArkaniHamed:1996jq}:
%\footnote{
%
%We redefine the gauge couplings $g_i$ as $1/(4g_i^2) - {k_i \left<Z\right>}/{M_P} \to 1/(4g_i^2)$.
%
%}
\begin{eqnarray}
M_1/M_2 = \frac{k_5 \mathcal{N}+ k_{1H}}{k_5} \frac{ g_{1H}^2}{g_5^2 + \mathcal{N} g_{1H}^2}, \ \ 
M_3/M_2=  \frac{k_5 + k_{3H}}{k_5} \frac{g_{3H}^2}{g_5^2 + g_{3H}^2},
\end{eqnarray}
where the real constant $\mathcal{N}$ depends on the $U(1)_H$ charge of the GUT breaking field.
%normalization of $U(1)_H$.
In the strong coupling limit, $g_{1H}^2, g_{3H}^2 \gg 1$,  the ratios of the gaugino masses are written in a simple form: $M_1/M_2 \simeq (k_5 \mathcal{N} + k_{1H})/(k_5 \mathcal{N})$ and $M_3/M_2 \simeq (k_5 + k_{3H})/k_5$. 

Together with Eqs.(\ref{eq:soft}) and (\ref{eq:ab-terms}), we have obtained a gaugino mediation model with non-zero soft masses of the Higgs sector and non-universal gaugino masses.
\\

%Note that if $H_u$ does not couple to $Z$, the soft mass for $m_{H_u}^2$ as well as $A_u$ vanishes. The Higgs B-term and trilinear couplings are given by
%\begin{eqnarray}
%B_\mu/\mu = A_d = A_e = 9k (1-f/3)^{-1/2} c_d \times m_{3/2}.
%\end{eqnarray}

%So far, we have neglected quantum corrections to soft masses.
%Quantum corrections by MSSM renormalizable couplings are taken into account by solving renormalization group equations of soft masses down to the soft mass scale.
Before closing this section, we discuss how large gravitino mass can be naturally taken in comparison with MSSM soft masses.
It is known that the gravitino as well as the SUSY breaking field cause various cosmological problems,
because they are easily produced in the early universe while  they are long-lived~\cite{Pagels:1981ke,Weinberg:1982zq,Khlopov:1984pf,Coughlan:1983ci}.
If the gravitino mass is large, these problems are relaxed because the gravitino and the SUSY breaking field are shorter-lived for larger masses.
It would be worth while to consider the possibility of a large gravitino mass in comparison with MSSM soft masses.

In order to evaluate the naturalness of a large gravitino mass, let us assume that couplings of the SUSY breaking field $Z$ with Higgs doublets ($\Delta K$), as well as that with gauginos ($k_5$, $k_{3H}$ and $k_{1H}$) are also absent and hence all soft masses vanish at the tree level.%
\footnote{Note that the $B_\mu$ term also vanishes in this limit in the gravitational SUSY breaking, as we have discussed. Thus, there is no so-called $B_\mu$ problem even in the sequestering limit.}
We estimate possible quantum corrections to soft masses under this assumption,
and require that the corrections do not upset our setup.

{ 
When MSSM soft masses vanish at the tree level, the MSSM
%with only renormalizable interactions
is a supersymmetric theory, where no soft masses are generated.%
\footnote{{This argument can be invalidated by couplings between regulator fields and the SUSY breaking field $Z$. See the comment in Appendix~\ref{sec:soft mass}.}}
Thus, quantum corrections to MSSM soft masses arise only from so-called the anomaly mediation~\cite{Randall:1998uk,Giudice:1998xp,Bagger:1999rd,D'Eramo:2012qd,Harigaya:2014sfa}
or from gravitational interactions~\cite{Antoniadis:1997ic,Luty:2002ff}.}%
\footnote{
{The anomaly mediation proportional to the VEV of the scalar auxiliary component of the supergravity multiplet is determined by the super-diffeomorphism~\cite{Harigaya:2014sfa} and hence cannot be eliminated.
In our setup, the VEV vanishes at the tree-level and hence that anomaly mediation is suppressed~\cite{Izawa:2010ym}.
}
}

%As is shown in Appendix~\ref{sec:soft mass}, the anomaly mediation vanishes, at the two loop level for scalar soft masses and the one-loop level for gaugino masses.
%This property comes from the vanishing VEV of the scalar auxiliary component of the supergravity multiplet in our  set up~\cite{IY}, and the sequestering between MSSM fields and the SUSY breaking field.

The possible largest quantum correction is the one-loop correction to scalar soft mass squared from gravitational interactions.
One-loop quantum corrections by gravitational interactions around the cut off scale
are expected to generate scalar soft mass squared as large as
\begin{eqnarray}
\Delta m_{\rm scalar}^2 \sim \frac{1}{16\pi^2} m_{3/2}^2 \left(\frac{\Lambda}{M_P}\right)^n,
%~~
%m_{\rm gaugino}, A_u, A_d, B \sim \frac{1}{16\pi^2} m_{3/2}\left(\frac{\Lambda}{M_*}\right)^n,
\end{eqnarray}
where $\Lambda$ and $n$ are
%the product of coupling constants,
the cut off of the theory and an integer, respectively.
%, the suppression scale of the higher-dimensional interactions, and the integer depending on the higher-dimensional interactions.
If $\Lambda \sim M_P$, the gravitino mass is at most $O(1)$ TeV for soft masses of $O(100)$ GeV.
{If $\Lambda \ll M_P$, however, as is assumed in the sequestering based on the brane world~\cite{Randall:1998uk}, the gravitino mass can be larger.
}

For some cases, 
one-loop corrections by gravitational interactions to scalar soft mass squared vanish~\cite{Binetruy:1987xj}, as is shown in Appendix~\ref{sec:soft mass}.
(There, we also discuss SUSY non-linear sigma models.)
%This is because corrections to scalar soft mass squared require both yukawa/gauge couplings and higher-dimensional interactions.
In that case, quantum corrections to scalar soft mass squared start from the two-loop level.
{Even if $\Lambda \sim M_P$,}
the two-loop corrections are at most comparable to possible one-loop corrections to gaugino masses, trilinear couplings and $B_\mu$ term.
Thus, the gravitino mass of $O(10)$ TeV is possible in this case, {even if $\Lambda \sim M_P$}.
%even if the gravitino mass is of $O(100)$ TeV, quantum corrections to soft masses are smaller than $O(1)$ TeV.
%The gravitino mass of $O(100)$ TeV is natural in this case.

%Finally, we comment on the possible origin of the suppression of soft masses ($c_u,d_u,c_d,d_d,$)

%Assuming $\Lambda \sim M_*$ to be conservative,
%%\footnote{If $\Lambda \ll M_*$, the constraint on the gravitino mass in Eq.~(\ref{eq:up on m32}) is relaxed.}
%possible quantum corrections are as large as %
%\begin{eqnarray}
%\Delta m_{\rm soft} \sim \frac{\lambda}{16\pi^2}m_{3/2},
%\end{eqnarray}
%%%
%where $\lambda$ is the product of coupling constant involved in the quantum corrections.
%Requiring that the tree-level soft masses $m_{\rm MSSM,soft}$ are smaller than this correction, we obtain the upper bound on the gravitino mass,
%%%
%\begin{eqnarray}
%m_{3/2} \lsim \frac{16\pi^2}{\lambda} m_{\rm MSSM,soft} \sim 100 {\rm TeV} \times \lambda^{-1} \frac{m_{\rm MSSM,soft}}{ 1 {\rm TeV}}.
%\label{eq:up on m32}
%\end{eqnarray}
%%%

\section{Higgs boson mass and $g-2$ of the muon}

In the MSSM, the observed Higgs boson mass at around 125 GeV is explained by a large stop mass and/or a large trilinear coupling of the stops~\cite{higgs_rad1}. These soft masses generate large radiative corrections to the Higgs potential: $\Delta m_{H_u}^2 \sim  (m_{\tilde{t}}^2$ or $A_t^2)$, where $m_{\tilde{t}}$ and $A_t$ are the stop mass and a trilinear coupling, respectively. For $m_{\tilde{t}}$, $A_t \sim 1\mathchar`-4$ TeV, the fine-tuning of parameters in the Higgs potential is required to explain the electroweak symmetry breaking (EWSB) scale. From EWSB conditions, $Z$ boson mass $m_Z$ is written as
%For large $\tan\beta$, the EWSB condition is written as
\begin{eqnarray}
\frac{m_Z^2}{2} &\simeq& - (m_{H_u}^2(M_{\rm GUT}) + \Delta m_{H_u}^2 + \mu^2) \nonumber \\
&&+ (m_{H_d}^2(M_{\rm GUT}) + \Delta m_{H_d}^2 - m_{H_u}^2(M_{\rm GUT}) - \Delta m_{H_u}^2)/ \tan^2\beta + \dots , \label{eq:ewsb}
\end{eqnarray}
where $\dots$ indicates terms suppressed by $1/\tan^n\beta$ $(n \geq 4)$.  Here, $m_{H_u}(M_{\rm GUT})$ and $m_{H_d}(M_{\rm GUT})$ denote the soft masses of the up- and down-type Higgs at the GUT scale, respectively, and $\Delta m_{H_u}^2$ and $\Delta m_{H_d}^2$ are the radiative corrections to $m_{H_u}^2$ and $m_{H_d}^2$.
To explain $m_Z \simeq 91.2$ GeV, $\Delta m_{H_u}^2 \sim (m_{\tilde{t}}^2$ or $A_t^2)$ needs to be cancelled by either $m_{H_u}^2 (M_{\rm GUT})$ or $\mu^2$.\footnote{
There is an exception where no large cancellation between $m_{H_u}^2 (M_{\rm GUT})$, $\Delta m_{H_u}^2$ and $\mu^2$ is required. For instance, in focus point gaugino mediation models~\cite{focus},  $\Delta m_{H_u}^2$ is small and the fine-tuning is significantly relaxed. 
}
Consequently, there arise two distinct regions: a small $\mu$ region and a large $\mu$ region. If the Higgs potential is tuned by the SUSY breaking parameter, $m_{H_u}^2 (M_{\rm GUT})$, it is likely that $\mu$ is of the order of $m_Z$. In contrast, if the EWSB scale is explained by the tuning of the SUSY invariant mass $\mu$, $\mu$ needs to be as large as $\mu \sim m_{\tilde{t}}$.

\paragraph{(i) small $\mu$ case}
When the Higgsino is light ($\mu$ is small), the SUSY contributions to the muon $g-2$ are dominated by the Wino-Higgsino-(muon-sneutrino) loop. In this case, the dominant SUSY contribution is given by~\cite{susy_gm2_2}
\begin{eqnarray}
(a_\mu)_{\tilde{W}-\tilde{H}-\tilde{\nu}_L} \simeq \frac{\alpha_2}{4\pi} \frac{m_\mu^2  (M_2 \mu) }{m_{{\tilde{\nu}}}^4}\tan\beta \cdot F_C\left(\frac{M_2^2}{m_{\tilde{\nu}}^2}, \frac{\mu^2}{m_{\tilde{\nu}}^2}\right) \ ,
\end{eqnarray}
where $F_C$ is a loop function (e.g.~$F_C(1,1)=1/2$). It is larger than the contribution from loops involving the Bino and smuons (see Eq.~(\ref{eq:g-2 large mu})).
%, which is dominant in the large $\mu$ case. 
%This is because the $SU(2)_L$ gauge coupling is larger than the $U(1)_Y$ gauge coupling;\footnote{
%Also, the loop function of the Wino-Higgsino-(muon-sneutrino) loop is larger than that of the Bino-smuons loop.
%}
Therefore, the small $\mu$ allows relatively heavier electroweakinos and sleptons to explain the muon $g-2$ than the large~$\mu$.

\begin{figure}[t]
\begin{flushleft}
\includegraphics[scale=1.00]{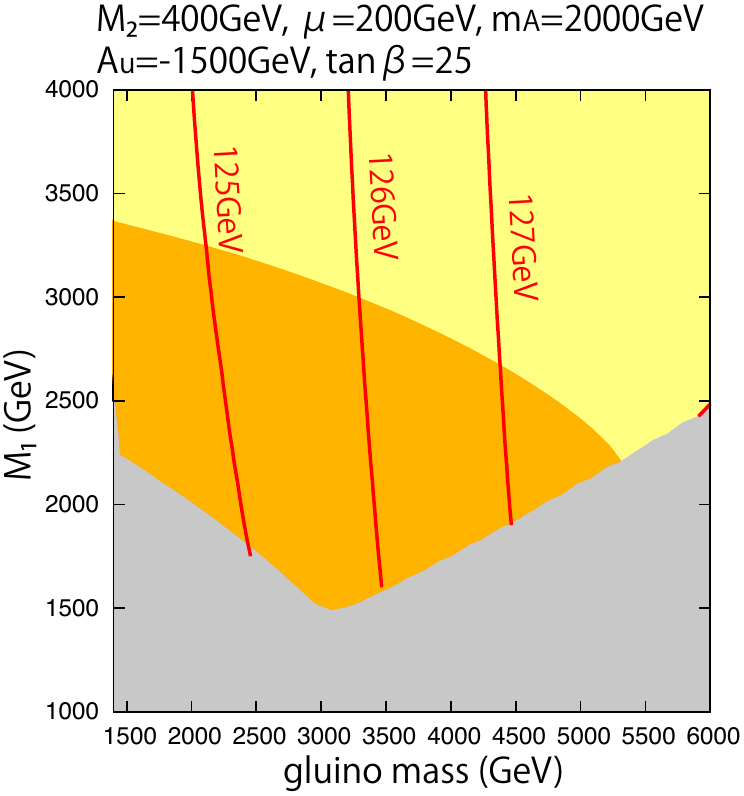}
\includegraphics[scale=1.00]{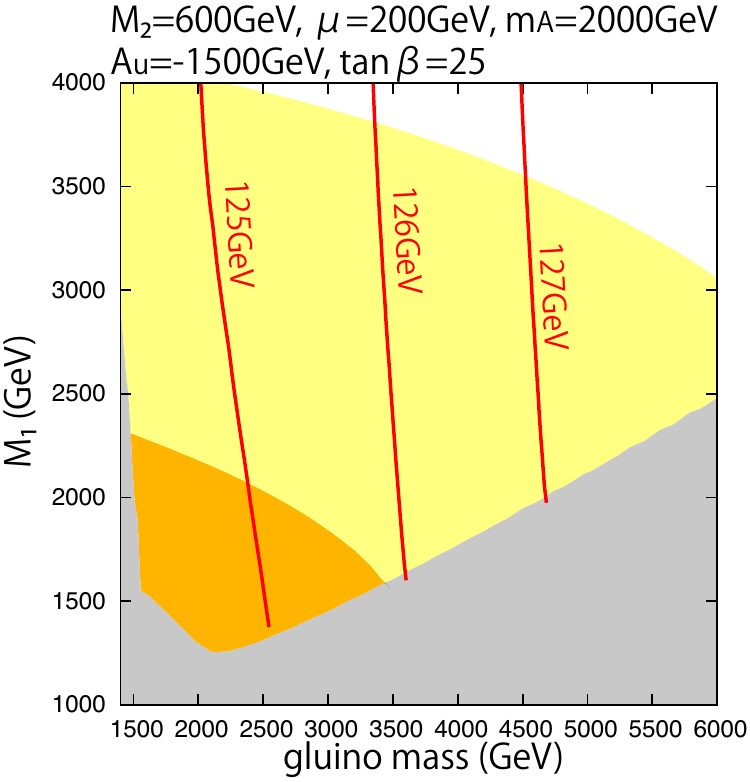}
\includegraphics[scale=1.00]{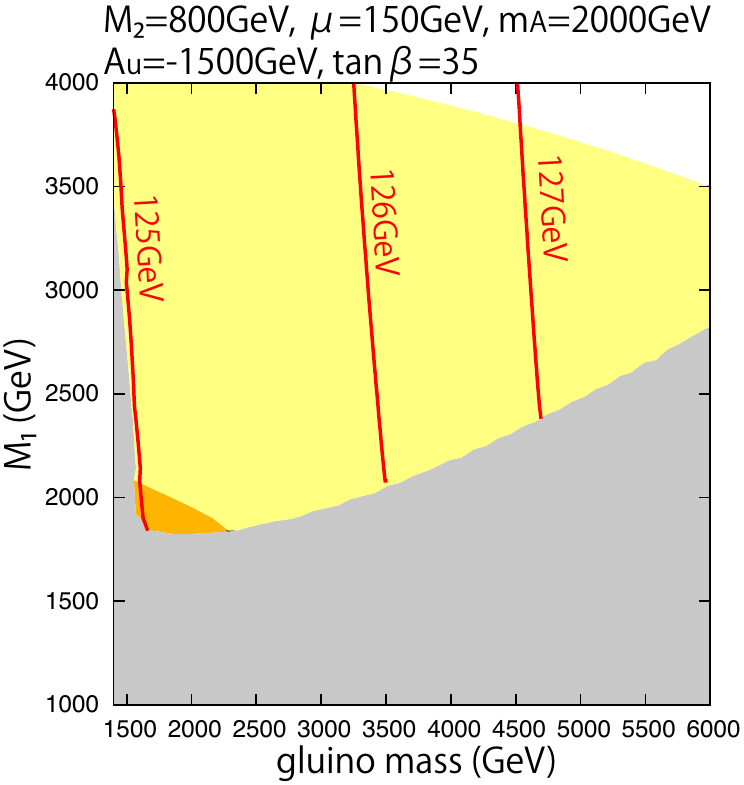}
\caption{The SUSY contribution to the muon $g-2$ and the Higgs boson mass (red solid line) in the small $\mu$ case.
In the orange (yellow) region, the muon $g-2$ is explained at 1(2) $\sigma$ level. In the top panels (the bottom panel), $\mu=200$ GeV\,(150 GeV) and $\tan\beta=20$\,(35). Here, $\alpha_S(m_Z)=0.1184$ and $m_t=173.3$ GeV.
}
\label{fig:gm2_1}
\end{flushleft}
\end{figure}

In Fig.~\ref{fig:gm2_1}, contours of the Higgs boson mass and the region consistent with the muon $g-2$ are shown. 
The mass of the CP-even Higgs boson is calculated by using {\tt FeynHiggs2.10.2}~\cite{feynhiggs} with the option resumming large logs. 
The SUSY contributions to the muon $g-2$, $\Delta a_\mu$, are also evaluated by {\tt FeynHiggs}.
The SUSY mass spectrum is evaluated by using {\tt Suspect} package~\cite{suspect} with modification suitable for our purpose.  In the top panels (the bottom panel), $B$, $m_{H_u}^2$ and $m_{H_d}^2$ at the GUT scale are taken such that $\tan\beta=25$ (35), $\mu=200$ GeV\,(150 GeV), and the physical mass of the CP-odd Higgs boson, $m_A=2000$ GeV, are reproduced.
The trilinear couplings are taken as $A_u=-1500{\rm GeV}, \ A_d=A_e=(B_\mu/\mu)-A_u$ (see Eq.(\ref{eq:ab-terms})).
The negative sign of $A_u$ at the GUT scale is taken so that contributions from gaugino loops are added to $A_u$ constructively.
%In the left panel, the trilinear coupling is assumed to be universal as $A_0=(B_\mu/\mu)/2$. 
%In the right panel, $A_u=-1500{\rm GeV}, \ \ A_d=A_e=(B_\mu/\mu)-A_u$. 
%However, the choice is not essential.
 We discard the gray region, since the stau becomes the lightest SUSY particle (LSP) or the stop is lighter than 600 GeV~\cite{stop_lhc}~\footnote{
In this region with the light stop, the constraint from the inclusive $b \to s \gamma$ decay is more stringent than the LHC stop searches.
 }; in the region of small $M_1$, the stau becomes light because of a small positive radiative correction from $M_1$.
The region consistent with the muon $g-2$ at 1$\sigma$ (2$\sigma$) level is shown in orange (yellow). 
Here, the deviation of the muon $g-2$ is evaluated as
\begin{eqnarray}
(a_\mu)_{\rm EXP} - (a_\mu )_{\rm SM} = (26.1 \pm 8.1) \cdot 10^{-10},
\end{eqnarray}
using the SM prediction in Ref.~\cite{gm2_sm1}.

One can see that the observed Higgs boson mass and muon $g-2$ are explained simultaneously for the physical gluino mass at around $2.5\mathchar`-3.0$ TeV and $M_2(M_{\rm GUT})=400\mathchar`-800$ GeV (corresponding to $M_2 (m_{\rm SUSY} )\simeq 290\mathchar`-620$\,GeV) (see also Table 1). 
Since squark masses are smaller than the physical gluino mass, the region consistent with the observed Higgs boson mass and the muon $g-2$ can be tested at the 14 TeV Large Hadron Collider (LHC), through productions of the gluino and squarks~\cite{ATLAS prospect 2013,ATLAS prospect 2014}.
%
%Here, $M_2(M_{\rm GUT}) = (400, 600, 800)$\,GeV corresponds to $M_2 (m_{\rm SUSY} )\simeq (290, 450, 620)$\,GeV. 
Note that the abundance of the lightest neutralino, which is mainly composed of the Higgsino with a mixture of the Wino, is much smaller than the observed value, $\Omega_{\rm DM} h^2 \simeq 0.12$~\cite{wmap, planck}, and hence, another candidate for dark matter is required.%
\footnote{
Since the spin-independent scattering cross section of this Higgsino-Wino neutralino is large as a few$\,\times 10^{-9}$ pb, the present model is excluded~\cite{LUX},
if the neutralino is a dominant component of dark matter.
}
% SI cross section is rather large as a few x 10^(-9) pb

In the region consistent with the muon $g-2$, the Wino has a small mass and hence the constraint from charino/neutralino searches at the LHC should be considered. The charged and neutral Winos produced by the electroweak interactions  decay into Higgsinos +  ($W$, $Z$, $h$), since left-handed sleptons are heavier than the Wino.
%The left-handed sleptons are heavy. Thus, the produced WIno decays into $W, Z$ and Higgs.} The charged and neutral Wino are produced by the electroweak interaction, and they subsequently decay into Higgsinos and  ($W$, $Z$, $h$). 
This process can be examined by the chargino/neutralino searches in the final state with two or three leptons and missing transverse momentum~\cite{atlas_2leptons, cms_electroweak}.
So far, the constraint is not very severe, and it is difficult to give a bound on the Wino mass for $\mu > 150$\,GeV; the region consistent with the muon $g-2$ is safe.
At the 14 TeV LHC, the Wino mass up to around 800 GeV can be excluded (discovered) for an integrated luminosity of 300 (3000) fb$^{-1}$~\cite{ATLAS prospect 2014}; therefore, it is expected that the region consistent with the muon $g-2$ at 1$\sigma$ level is tested at the 14 TeV LHC.

\paragraph{(ii) large $\mu$ case}
If the Higgsino is heavy, the Bino-(L-smuon)-(R-smuon) loop dominates the SUSY contributions to the muon $g-2$. The contribution from this loop is given by~\cite{susy_gm2_2}
\begin{eqnarray}
\label{eq:g-2 large mu}
(a_\mu)_{\tilde{B}-\tilde{\mu}_L-\tilde{\mu}_R} \simeq \frac{3}{5} \frac{\alpha_1}{4\pi} \frac{m_\mu^2 \mu}{M_1^3}  \tan\beta \cdot F_N\left(\frac{m_{\tilde{\mu}_R}^2}{M_1^2}, \frac{m_{\tilde{\mu}_L}^2}{M_1^2}\right),
\end{eqnarray}
where $F_N$ is a loop function (e.g.~$F_N(1,1)=1/6$).
%Since this contribution usually smaller than that of the Wino-Higgsino, it tends to be excluded by the electroweak SUSY searches at the LHC.
In this region, a small Bino mass is required to enhance $(a_\mu)_{\tilde{B}-\tilde{\mu}_L-\tilde{\mu}_R}$. As a result,   the stau tends to be tachyonic in a gaugino mediation model.
This tachyonic stau is avoided by the positive contribution from $(m_{H_d}^2-m_{H_u}^2)$ through the renormalization group evolution. Indeed, the one-loop renormalization group equation for the right handed slepton is given by
\begin{eqnarray}
\frac{d m_{E^c}^2}{d t} \ni \frac{1}{16\pi^2} \Bigl[ - \frac{6}{5} g_1^2 (m_{H_d}^2 - m_{H_u}^2) \Bigr].
\end{eqnarray}
This contribution  vanishes if $m_{H_u}^2=m_{H_d}^2$. However, it is sizable when $m_{H_d}^2 \gg m_{H_u}^2$. In our gaugino mediation model, $H_u$ and $H_d$ couple to the SUSY breaking field $Z$; therefore, $(m_{H_d}^2-m_{H_u}^2)$ can be positive at the high energy scale resulting in the positive mass squared of the stau.

\begin{figure}[t]
\begin{flushleft}
\includegraphics[scale=1.00]{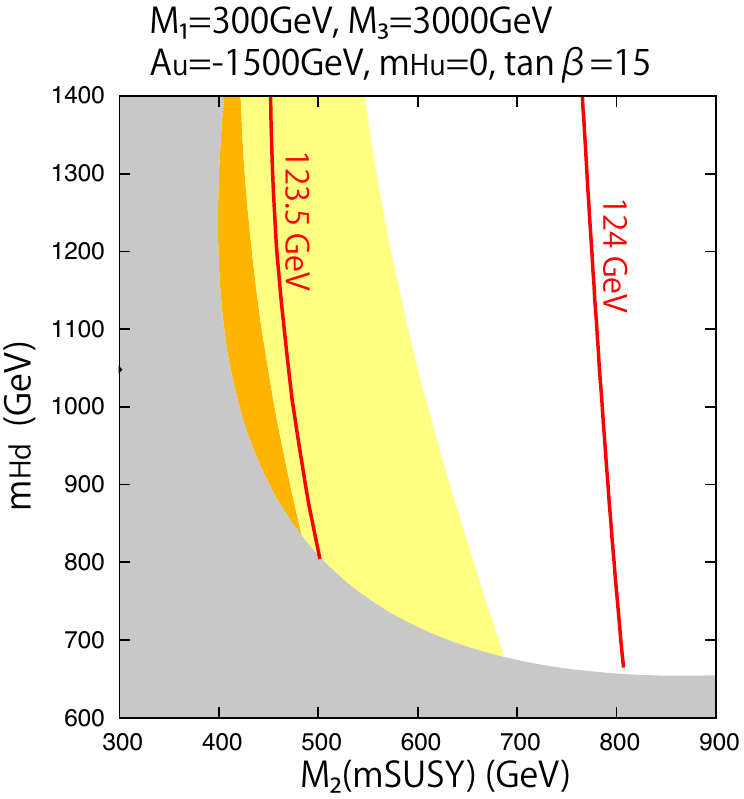}
\includegraphics[scale=1.00]{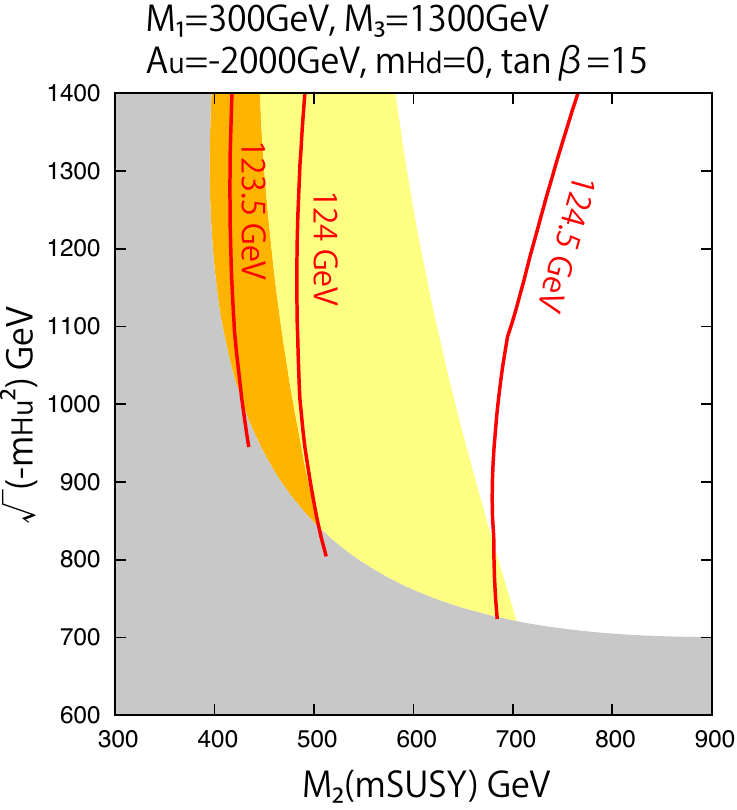}
\includegraphics[scale=1.00]{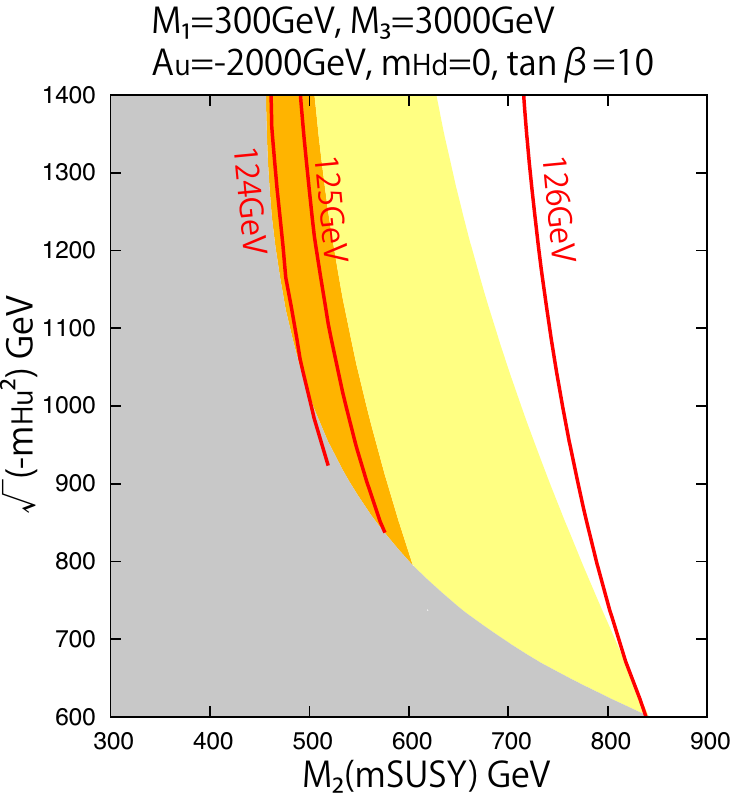}
\caption{The SUSY contribution to the muon $g-2$ and the Higgs boson mass in the large $\mu$ case. The different  values of $M_3$, $M_3=1300$ GeV (top panels) and 3000 GeV (bottom panel) are taken. 
In the orange (yellow) region, the muon $g-2$ is explained at 1(2) $\sigma$ level.
In the top-left panel, $m_{H_u}=0$, while in the other panels, $m_{H_d}=0$ ).
}
\label{fig:gm2_2}
\end{flushleft}
\end{figure}

In Fig.~\ref{fig:gm2_2}, we show $\Delta a_{\mu}$ and the contours of $m_h$ in the large  $\mu$ cases.  In the top-left panel, $m_{H_d}^2(M_{\rm GUT}) > 0$ and $m_{H_u}^2=0$ are taken, and in other two panels, $m_{H_d}^2(M_{\rm GUT}) =0$ and $m_{H_u}^2(M_{\rm GUT}) < 0$ are  taken.
 The trilinear coupling $A_u$ is $A_u=-1500$\,GeV in the top-left panel, and $A_u=-2000$\,GeV in the other two panels. Other trilinear couplings satisfy the condition in Eq.(\ref{eq:ab-terms}), $A_d=A_e=(B_\mu/\mu)-A_u$. 
The gray region is exclude since the stau becomes the lightest SUSY particle.
 % (LSP).
On the edge of the stau LSP region, i.e., for $m_{\chi_1^0} \sim m_{\tilde{\tau}_1}$, the relic abundance of the neturalino is consistent with the observed value, $\Omega_{\rm DM} h^2 \simeq 0.12$~\cite{wmap,planck}. This is because the coannihilation~\cite{Griest:1990kh} with the lighter stau reduces the relic abundance of the lightest neutralino efficiently.

 It is shown that the muon $g-2$ is explained at 1$\sigma$ level for $M_2(m_{\rm SUSY}) < 500\, (600)$\,GeV and the gluino mass of 2.8 (6.1)\,TeV. The calculated Higgs boson mass can be consistent with the observed mass at around 125 GeV by taking into account the uncertainties from the top pole mass and theoretical calculation of the Higgs boson mass. However, as shown below, this region is rather severely constrained from the chargino/neuralino searches at the LHC. 
Here, $M_3=(1300, 3000)$\,GeV corresponds to $m_{\rm gluino}=(2.8, 6.1)$\,TeV and $m_{\rm squark}=(2.4, 5.2)$\,TeV, where $m_{\rm gluino}$ and $m_{\rm squark}$ are the physical gluino mass and squark mass, respectively.

%\paragraph{chargino/slepton search}
In contrast to the small $\mu$ case, the region consistent with the muon $g-2$ in the large $\mu$ case is rather severely constrained by chargino/neutralino searches at the LHC. This is because the left-handed sleptons are lighter than the Wino in this region, and hence the Wino can decay into an on-shell slepton and a lepton, 
with the slepton decaying into a neutralino and a lepton.
As a result, this region is severely constrained by searches for the electroweak production of the chargino and neutralino in a final state with three leptons and missing transverse momentum~\cite{cms_electroweak, atlas_3leptons}. From these searches, the Wino mass  is constrained to be larger than $600\mathchar`-700$\,GeV, depending on the slepton mass. Considering this constraint, the muon $g-2$ is explained at $1.5\mathchar`-2$$\sigma$ ($1\mathchar`-1.5$$\sigma$) level for $M_3=1300~(3000)$\,GeV (see also Table~\ref{table:spectrum2}).
For larger $M_3$, e.g., $M_3=5200$ GeV,
the muon $g-2$ is explain at 1$\sigma$ level with a chargino mass
larger than $700$ GeV (see P6 in Table~\ref{table:spectrum2}).
Note that a large gluino mass indirectly enhances $(a_\mu)_{\tilde{B}-\tilde{\mu}_L-\tilde{\mu}_R}$ for fixed $\tan\beta$, bino and smuon masses. Large $M_3$ generates large $|m_{H_u}^2|$ through radiative corrections at the two-loop level. With this large $|m_{H_u}^2|$, $\mu$ is determined to be large from the EWSB condition in Eq.~(\ref{eq:ewsb}). Consequently, $(a_\mu)_{\tilde{B}-\tilde{\mu}_L-\tilde{\mu}_R}$ proportional to $\mu$ is enhanced.\footnote{
Large $\mu \tan\beta$ generates to a charge breaking minimum in the Higgs-stau potential, which can be deeper than the EWSB minimum. Therefore, the size of $\mu\tan\beta$ is constrained from above by the stability of the EWSB minimum~\cite{charge_breaking}.
}
%A heavier gluino allows a  heavier Wino, and it becomes easier to avoid the constraint from the chargino/neutralino searches at the LHC.
This is the reason why favored gluino masses are relatively larger than those in the small $\mu$ case.

%\paragraph{Some table of the mass spectrum}
Finally, some mass spectra for small $\mu$  and large $\mu$ are shown in Table \ref{table:spectrum1} and \ref{table:spectrum2}. In small $\mu$ cases, the branching ratio of the inclusive $b \to s \gamma$ decay is enhanced for  large $A_t$; we calculate $\Delta  {\rm Br}(b \to s \gamma)= {\rm Br}(b \to s \gamma)_{\rm MSSM}-{\rm Br}(b \to s \gamma)_{\rm SM}$ in P1$-$P2 by using {\tt SuperIso} package~\cite{superiso}, and demand that $-0.34\cdot 10^{-4}  < \Delta  {\rm Br}(b \to s \gamma) < 0.90 \cdot 10^{-4}$. Here, the  required range of $\Delta  {\rm Br}(b \to s \gamma)$ is given by the difference between the experimental value of ${\rm Br}(b \to s \gamma)$~\cite{HFAG} and SM prediction~\cite{bsg_sm} with an inclusion of 2$\sigma$ error. 

%{\bf In large $\mu$ cases, the left-right mixing of the stau, $m_{\tau} \mu\tan\beta$, is large. This left-right mixing induces  a charge breaking minimum, which can be deeper than the EWSB minimum~\cite{Rattazzi:1996fb}. Therefore, we require that the life-time of the EWSB minimum be longer than the age of the universe in P4$-$P6. 
%}
%%%%%%%%%%%%%%%
%%%%%%%%%%%%%%%

%%%%%%%%%%%%%%%
%%%%%%%%%%%%%%%

%%%%%%%%%%%%%%%%%%%%%%%%%%%%%%%%%%%%%%%%%%%%%
\begin{table}[t!]
  \begin{center}
    \begin{tabular}{  c | c  }
            P1 & \\
\hline
    $M_1(M_{\rm GUT})$ & 2200 GeV \\
    $M_2(M_{\rm GUT})$ & 400 GeV \\
    $M_3(M_{\rm GUT})$ & 1100 GeV \\
    $A_u(M_{\rm GUT})$ & -1300 GeV\\
    $\tan \beta$ & 20 \\
    $\mu$ & 200 GeV \\
    $m_A$ & 2 TeV \\
    \hline
\hline    
    $m_{\rm gluino}$ & 2.4 TeV \\
      $m_{\tilde{q}}$ & 2.1 TeV \\
    $m_{\tilde{t}_{1,2}}$ & 1.4, 1.8 TeV \\
    $m_{\tilde{e}_L} (m_{\tilde{\mu}_L})$ & 450 GeV\\
    $m_{\tilde{e}_R} (m_{\tilde{\mu}_R})$ & 836 GeV\\
    $m_{\tilde{\tau}_1}$ & 361 GeV\\
     $m_{\chi_1^0}$, $m_{\chi_2^0}$ & 179, 210 GeV \\
     $m_{\chi_3^0}$, $m_{\chi_4^0}$ & 342, 935 GeV \\
     $m_{\chi_1^{\pm}}$, $m_{\chi_2^{\pm}}$ & 184, 343 GeV \\
     $m_{h}$ & 124.5 GeV \\
    $ \Delta a_\mu$ & $2.20 \times 10^{-9}$ \\
%    $ \Delta {\rm Br}(b \to X_s \gamma)$ & $-2.85 \times 10^{-5}$\\  	
     $ \Delta {\rm Br}(b \to s \gamma)$ & $-2.9 \times 10^{-5}$\\  	
    \end{tabular}
        \hspace{20pt}
%%%%%%%%%%%%%%
        \begin{tabular}{  c | c  }
                    P2 & \\
\hline
    $M_1(M_{\rm GUT})$ & 1900 GeV \\
    $M_2(M_{\rm GUT})$ & 740 GeV \\
    $M_3(M_{\rm GUT})$ & 1400 GeV \\
    $A_u(M_{\rm GUT})$ & -1300 GeV\\
    $\tan \beta$ & 35 \\
    $\mu$ & 150 GeV \\
    $m_A$ & 2 TeV \\
    \hline
\hline    
    $m_{\rm gluino}$ & 3.0 TeV \\
      $m_{\tilde{q}}$ & 2.6 TeV \\
    $m_{\tilde{t}_{1,2}}$ & 1.8, 2.2 TeV \\
    $m_{\tilde{e}_L} (m_{\tilde{\mu}_L})$ & 573 GeV\\
    $m_{\tilde{e}_R} (m_{\tilde{\mu}_R})$ & 721 GeV\\
    $m_{\tilde{\tau}_1}$ & 174 GeV\\
     $m_{\chi_1^0}$, $m_{\chi_2^0}$ & 145, 159 GeV \\
     $m_{\chi_3^0}$, $m_{\chi_4^0}$ & 602, 806 GeV \\
     $m_{\chi_1^{\pm}}$, $m_{\chi_2^{\pm}}$ & 151, 602 GeV \\
     $m_{h}$ & 125.3 GeV \\
    $ \Delta a_\mu$ & $1.97 \times 10^{-9}$ \\
%    $ \Delta {\rm Br}(b \to X_s \gamma)$ & $-2.48 \times 10^{-5}$\\  	
    $ \Delta {\rm Br}(b \to s \gamma)$ & $-2.5 \times 10^{-5}$\\  	
    \end{tabular}
    \caption{The mass spectra with the small $\mu$. Here, $\Delta  {\rm Br}(b \to s \gamma)= {\rm Br}(b \to s \gamma)_{\rm MSSM}-{\rm Br}(b \to s \gamma)_{\rm SM}$.
   }
  \label{table:spectrum1}
  \end{center}
\end{table}
%%%%%%%%%%%%%%%%%%%%%%%%%%%%%%%%%%%%%%%%%%%%%

%%%%%%%%%%%%%%%%%%%%%%%%%%%%%%%%%%%%%%%%%%%%%
\begin{table}[t!]
  \begin{center}
      \begin{tabular}{  c | c  }
    P3 & \\
\hline
    $M_1(M_{\rm GUT})$ & 300 GeV \\
    $M_2(M_{\rm GUT})$ & 760 GeV \\
    $M_3(M_{\rm GUT})$ & 1300 GeV \\
    $A_u(M_{\rm GUT})$ & -2000 GeV\\
    $\tan \beta$ & 15 \\
    $m_{H_u}^2(M_{\rm GUT})$ & $-6 \cdot 10^5$\,GeV$^2$ \\
    $m_{H_d}^2(M_{\rm GUT})$ & 0 \\
    \hline
\hline    
    $m_{\rm gluino}$ & 2.8 TeV \\
      $m_{\tilde{q}}$ & 2.5 TeV \\
    $m_{\tilde{t}_{1,2}}$ & 1.9, 2.2 TeV \\
    $m_{\tilde{e}_L} (m_{\tilde{\mu}_L})$ & 471 GeV\\
    $m_{\tilde{e}_R} (m_{\tilde{\mu}_R})$ & 212 GeV\\
    $m_{\tilde{\tau}_1}$ & 120 GeV\\
     $m_{\chi_1^0}$, $m_{\chi_2^0}$ & 118, 609 GeV \\
 %    $m_{\chi_3^0}$, $m_{\chi_4^0}$ & 602, 806 GeV \\
     $m_{\chi_1^{\pm}}$, $m_{\chi_2^{\pm}}$ & 609, 2006 GeV \\
     $m_{h}$ & 124.3 GeV \\
    $ \Delta a_\mu$ & $1.40 \times 10^{-9}$ \\
    \end{tabular}
    \hspace{20pt}
 %%%%%%%%%%%%%%%%
 %
    \begin{tabular}{  c | c  }
        P4 & \\
\hline
    $M_1(M_{\rm GUT})$ & 300 GeV \\
    $M_2(M_{\rm GUT})$ & 780 GeV \\
    $M_3(M_{\rm GUT})$ & 3000 GeV \\
    $A_u(M_{\rm GUT})$ & -2000 GeV\\
    $\tan \beta$ & 10 \\
    $m_{H_u}^2(M_{\rm GUT})$ & $-7.6\cdot10^5$\,GeV$^2$ \\
    $m_{H_d}^2(M_{\rm GUT})$ & $0$ \\
    \hline
\hline    
    $m_{\rm gluino}$ & 6.1 TeV \\
      $m_{\tilde{q}}$ & 5.2 TeV \\
    $m_{\tilde{t}_{1,2}}$ & 4.4, 4.9 TeV \\
    $m_{\tilde{e}_L} (m_{\tilde{\mu}_L})$ & 423 GeV\\
    $m_{\tilde{e}_R} (m_{\tilde{\mu}_R})$ & 218 GeV\\
    $m_{\tilde{\tau}_1}$ & 118 GeV\\
     $m_{\chi_1^0}$, $m_{\chi_2^0}$ & 107, 606 GeV \\
%     $m_{\chi_3^0}$, $m_{\chi_4^0}$ & 602, 806 GeV \\
     $m_{\chi_1^{\pm}}$, $m_{\chi_2^{\pm}}$ & 606, 3671 GeV \\
     $m_{h}$ & 125.2 GeV \\
    $ \Delta a_\mu$ & $1.88 \times 10^{-9}$ \\
    \end{tabular}
      \end{center}
    %%%%%%%%%%%%%
    \hspace{50pt}   
     \begin{tabular}{  c | c  }
        P5 & \\
\hline
    $M_1(M_{\rm GUT})$ & 300 GeV \\
    $M_2(M_{\rm GUT})$ & 900 GeV \\
    $M_3(M_{\rm GUT})$ & 3000 GeV \\
    $A_u(M_{\rm GUT})$ & -2000 GeV\\
    $\tan \beta$ & 10 \\
    $m_{H_u}^2(M_{\rm GUT})$ & 0 \\
    $m_{H_d}^2(M_{\rm GUT})$ & $6\cdot 10^5$\,GeV$^2$ \\
    \hline
\hline    
    $m_{\rm gluino}$ & 6.1 TeV \\
      $m_{\tilde{q}}$ & 5.2 TeV \\
    $m_{\tilde{t}_{1,2}}$ & 4.4, 4.9 TeV \\
    $m_{\tilde{e}_L} (m_{\tilde{\mu}_L})$ & 515 GeV\\
    $m_{\tilde{e}_R} (m_{\tilde{\mu}_R})$ & 203 GeV\\
    $m_{\tilde{\tau}_1}$ & 113 GeV\\
     $m_{\chi_1^0}$, $m_{\chi_2^0}$ & 107, 707 GeV \\
%     $m_{\chi_3^0}$, $m_{\chi_4^0}$ & 602, 806 GeV \\
     $m_{\chi_1^{\pm}}$, $m_{\chi_2^{\pm}}$ & 707, 3602 GeV \\
     $m_{h}$ & 125.7 GeV \\
    $ \Delta a_\mu$ & $1.41 \times 10^{-9}$ \\
    \end{tabular}
        \hspace{20pt}   
         \begin{tabular}{  c | c  }
        P6 & \\
\hline
    $M_1(M_{\rm GUT})$ & 300 GeV \\
    $M_2(M_{\rm GUT})$ & 940 GeV \\
    $M_3(M_{\rm GUT})$ & 5200 GeV \\
    $A_u(M_{\rm GUT})$ & -2000 GeV\\
    $\tan \beta$ & 7 \\
    $m_{H_u}^2(M_{\rm GUT})$ & $-1.2\cdot 10^6$\,GeV$^2$ \\
    $m_{H_d}^2(M_{\rm GUT})$ & 0 \\
    \hline
\hline    
    $m_{\rm gluino}$ & 10.3 TeV \\
      $m_{\tilde{q}}$ & 8.7 TeV \\
    $m_{\tilde{t}_{1,2}}$ & 7.4, 8.1 TeV \\
    $m_{\tilde{e}_L} (m_{\tilde{\mu}_L})$ & 419 GeV\\
    $m_{\tilde{e}_R} (m_{\tilde{\mu}_R})$ & 231 GeV\\
    $m_{\tilde{\tau}_1}$ & 104 GeV\\
     $m_{\chi_1^0}$, $m_{\chi_2^0}$ & 92, 708 GeV \\
%     $m_{\chi_3^0}$, $m_{\chi_4^0}$ & 602, 806 GeV \\
     $m_{\chi_1^{\pm}}$, $m_{\chi_2^{\pm}}$ & 708, 5828 GeV \\
     $m_{h}$ & 124.8 GeV \\
    $ \Delta a_\mu$ & $1.86 \times 10^{-9}$ \\
    \end{tabular}
    \caption{The mass spectra with the large $\mu$. Here, $\mu \simeq m_{\chi_2^{\pm}}$.
   }
  \label{table:spectrum2}

\end{table}
%%%%%%%%%%%%%%%%%%%%%%%%%%%%%%%%%%%%%%%%%%%%%

\section{Conclusion and discussion}
We have shown that  the observed Higgs boson mass at around 125 GeV and the anomaly of the muon $g-2$ are explained simultaneously in our gaugino mediation models. There is no SUSY CP problem thanks to the shift symmetry of the SUSY breaking field $Z$, and the gravitational SUSY breaking mechanism.
The Higgs doublets are assumed to couple to $Z$, giving non-zero soft masses of the Higgs doublets and the trilinear coupling of the stops. 
With this trilinear coupling, the Higgs boson mass of 125 GeV is explained relatively easily: colored SUSY particles can be lighter than 3\,TeV, and they are expected to be produced at the 14 TeV LHC. 
Thanks to the non-zero soft masses of the Higgs doublets, $\mu$ parameter can be small if the Higgs potential is tuned by these soft masses. In this case, the muon $g-2$ is explained at $1\sigma$ level avoiding the constraints from the chargino/neutralino searches.
With the small $\mu$, light Higgsinos as well as sleptons are targets of searches at lepton colliders such as the
International Linear Collider experiments.

The muon $g-2$ is consistent with the large $\mu$ case as well. In the large $\mu$ case, the Higgs potential is tuned by this $\mu$ parameter rather than the soft masses of the Higgs doublets. Compared to the small $\mu$ case, a relatively heavy gluino is favored for the muon $g-2$, and the gluino and squarks are too heavy to be tested even at the 14 TeV LHC. However, this region can be easily covered through the chargino/slepton searches at the LHC~\cite{Iwamoto:2014ywa}.

Finally, let us comment on possible problems concerning cosmological aspects; the gravitino problem and the Polonyi problem. By allowing small couplings between $Z$ and Higgs doublets, the gravitino mass can be larger than $O(10)$ TeV. In this case, the cosmological gravitino problem is significantly relaxed, compared to the case with the gravitino mass of $O(1)$ TeV.
The Polony problem caused by $Z$ can be solved by the adiabatic solution provided that $Z$ strongly couples to an inflaton~\cite{linde, nty2011}. Or maybe the Polonyi problem is simply absent due to an anthropic  reason;
 note that the constraint from the Big Bang Nucleosynthesis, which would not be avoided by an anthropic reason,
is avoided for a sufficiently large mass of $Z$, $m_Z >O(10)$ TeV, because $Z$ decays before the Big Bang Nucleosynthesis starts.

\section*{Acknowledgments}
We thank Sho Iwamoto for useful discussion.
This work is supported by Grant-in-Aid for Scientific research from the
Ministry of Education, Science, Sports, and Culture (MEXT), Japan,
No.\ 26104009 and 26287039 (T.\,T.\,Y.),
and also by World Premier International Research Center Initiative (WPI Initiative), MEXT, Japan (K.\,H.~and T.\,T.\,Y.).
The work of K.\,H.\, is supported in part by a JSPS Research Fellowships for Young Scientists.

\appendix
\section{Soft scalar masses}
\label{sec:soft mass}
In this appendix, we calculate soft scalar masses of MSSM particles shown in Sec.~\ref{sec:gaugino med}.
For simplicity, we put $M_P = 1$ in this appendix.

\subsection{Tree level soft scalar masses}

In general, the scalar potential is given by the Kahler potential $K(X^I, X^{\dag\bar{J}})$ and the super potential $W(X^I)$ as%
\footnote{We omit the $D$ term potential, which is irrelevant for our discussion.}
\begin{eqnarray}
V = e^K \left[ 
K^{I\bar{J}} (W_I + K_I W) (W_J + K_J W)^\dag - 3 |W|^2
\right].
\end{eqnarray}
Here, $I,J,\cdots$ and $\bar{I}, \bar{J},\cdots$ indicate chiral and anti-chiral fields, respectively.
Lower indices denote derivatives with respect to the corresponding field.
$K^{I\bar{J}}$ is the inverse of the matrix $K_{I\bar{J}}$.

Let us first consider the following Kahler potential and the super potential,
\begin{eqnarray}
K = g (Z + Z^\dag) + h_i (Z + Z^\dag)Q_i^\dag Q_i,\\
W = \mathcal{C} + \tilde{y} Q_1 Q_2 Q_3 + \tilde{\mu} Q_4 Q_5,
\end{eqnarray}
where $Q_i~(i=1\mathchar`-5)$, $\tilde{y}$ and $\tilde{\mu}$ are matter chiral fields, the yukawa coupling, and the mass term, respectively. $g$ and $h_i$ are real functions.
Note that fields $Q_i$ are not canonically normalized.
Canonically normalized fields are given by
\begin{eqnarray}
Q_i^c = h_i^{1/2} Q_i.
\label{eq:canonical}
\end{eqnarray}
Then the yukawa coupling and the mass term for the canonically normalized fields are
\begin{eqnarray}
y = e^{g/2} (h_1 h_2 h_3)^{-1/2} \tilde{y} ,~~\mu = e^{g/2} (h_4 h_5)^{-1/2} \tilde{\mu}.
\end{eqnarray}

The potential of $Z$ is given by
\begin{eqnarray}
V(Z,Z^\dag) = e^g |\mathcal{C}|^2\left[
g^{''-1} g^{'2} -3
\right].
\end{eqnarray}
Vanishing of the cosmological constant requires that
\begin{eqnarray}
g^{'2} = 3 g''
\label{eq:CC cancel}
\end{eqnarray}
at the vacuum.

Scalar soft masses are given by
\begin{eqnarray} 
V_{\rm soft} &=&  m_i^2 |Q_i^{c}|^2 + \left [ y A Q_1^c Q_2^c Q_3^c + B_\mu Q_4^c Q_5^c + {\rm h.c.} \right], \nonumber \\
m_i ^2 &=& \left[
 1 - 9 (g')^{-2} \left({\rm ln} h_i\right)''
\right]\times m_{3/2}^2 , \nonumber \\
A &=& \sum_{i=1,2,3} \left(
1 - \frac{3}{g'} \left({\rm ln}h_i\right)' 
\right) \times m_{3/2} , \nonumber \\
B_\mu / \mu &=&\sum_{i=4,5} \left(
1 - \frac{3}{g'} \left({\rm ln}h_i\right)' 
\right) \times m_{3/2},
\label{eq:soft mass general}
\end{eqnarray}
where primes denote the derivative with respect to $Z + Z^\dag$.
Here, we have used $e^{K/2} \mathcal{C^*} = m_{3/2}$ and Eq.~(\ref{eq:CC cancel}).

The Kahler potential discussed in Sec.~\ref{sec:gaugino med} corresponds to the case with
\begin{eqnarray}
g &=& -3 {\rm ln} \left( 1 - f/3\right), \nonumber\\
h_i &=& 
\left\{
\begin{array}{ll}
\left(1-f/3\right)^{-1} \left( 1+ c_u\left( Z + Z^\dag \right) + d_u \left( Z + Z^\dag \right)^2  \right) & {\rm for~}Q_i = H_u, \\
\left(1-f/3\right)^{-1} \left( 1+ c_d\left( Z + Z^\dag \right) + d_d \left( Z + Z^\dag \right)^2  \right) & {\rm for~}Q_i = H_d, \\
\left(1-f/3\right)^{-1} & {\rm for~others}.
\end{array}
\right.
\label{eq:correspondence}
\end{eqnarray}
From Eqs.~(\ref{eq:soft mass general}) and (\ref{eq:correspondence}), we obtain soft masses shown in Eqs.~(\ref{eq:soft}) and (\ref{eq:ab-terms}).
Scalar soft mass squared of squarks and sleptons vanish.

\subsection{Quantum corrections to soft scalar mass squared}

%So far, we have discussed tree-level soft masses.
We are interested in the case where the gravitino mass is far larger than soft masses of MSSM,
because the gravitino problem and the Polonyi problem are relaxed in this case.
Let us assume vanishing  soft masses at the tree-level (in the sequestering discussed in the main text, $\Delta K =0$ in Eq.~(\ref{eq:deltaK}) and $k_5 = k_{3H} = k_{1H}=0$ in Eq.~(\ref{eq:gauge kin})),
and discuss how large quantum corrections to soft masses are expected.

As we have discussed in the main text, when MSSM soft masses vanish at the tree-level, renormalizable interactions in MSSM does not generate soft masses.
Then quantum corrections to MSSM soft masses originate only from the anomaly mediation or from higher-dimensional interactions such as gravitational interactions.
The possible largest correction is the one-loop correction to scalar soft mass squared from higher-dimensional interactions,
which we investigate here.
%In the following, we show that the one-loop correction is absent in our set up.

MSSM field couples to the SUSY breaking field and the gravitino through higher dimensional interactions.
These interactions are expected to generate MSSM soft masses through quantum corrections.
We treat the quantum corrections by introducing a cut off $\Lambda$ to the theory, and
regard the action with vanishing MSSM soft masses as a Wilsonian action at the cut off scale.
Then quantum corrections to scalar soft mass squared are given by the following form,
\begin{eqnarray}
\Delta m_{\rm scalar}^2 \sim \frac{1}{(16\pi^2)^m} m_{3/2}^2 \left( \frac{\Lambda}{M_*} \right)^n,
\end{eqnarray}
where $M_*$ is the suppression scale of higher dimensional interactions such as the Planck scale, $m$ is the number of loops, and $n$ is an integer.
% depending on higher dimensional interactions.

In the following, we show that one-loop corrections to soft mass squared are absent,
% in our set up.
when the tree-level Kahler potential is either of the following forms;
%Let us consider the following forms of the Kahler potential;
%%
\begin{eqnarray}
\label{eq:noscale}
K &=& K (Z +Z^\dag + Q_i^\dag Q_i), \\
\label{eq:cpn}
K &=& K \left({\rm ln}\left(1 + Q_i^\dag Q_i\right) + Z + Z^\dag  \right).
\end{eqnarray}
The former is the case with the sequestered form in Eq.~(\ref{eq:sequestering}) with $f(Z + Z^\dag)$ linear in $Z + Z^\dag$, so-called the no-scale structure.%
\footnote{For the no-scale case, the tree-level potential of $Z$ also vanishes. The potential is given by one-loop corrections and hence the scalar component of $Z$ is lighter than the gravitino.}
The latter is the case with the SUSY $CP^N\simeq SU(N+1) / SU(N)\times U(1)$ nonlinear sigma model, where $N$ is the number of chiral superfields $Q_i$.
The proof is parallel to the one given in Ref.~\cite{Binetruy:1987xj}.

In the limit of vanishing yukawa couplings and gauge couplings, the action is invariant under the following  transformation of superfields;
\begin{eqnarray}
\delta Q_i = \epsilon_i,~~ \delta Z = \epsilon_i^* Q_i & :{\rm for~Eq.}~(\ref{eq:noscale}),\\
\delta Q_i = 
- \frac{i}{2}\theta_{i1}(1-Q_i^2) + \frac{1}{2}\theta_{i2} (1 + Q_i^2) - i \theta_{i3} Q_i, &\nonumber \\
\delta Z = 
- \frac{i}{2} \theta_{i1} Q_i - \frac{1}{2}\theta_{i2} Q_i + \frac{i}{2}\theta_{i3}
 & :{\rm for~ Eq.}~(\ref{eq:cpn}),
\end{eqnarray}
where $\epsilon_i$ and $\theta_i$ are complex and real infinitesimal parameters of transformations.
The latter transformation is nothing but $SU(2)$ subgroups of $SU(N+1)$.

In both cases, quantum corrections without yukawa nor gauge coupling constants can modify the Kahler potential only to the following form consistent with the symmetries;
%%%
\begin{eqnarray}
K = F(Z + Z^\dag + Q_i^\dag Q_i) + \cdots ,
\label{eq:Kahler correction}
\end{eqnarray}
%%%
where $F$ is some real function and $\cdots$ denote terms higher order in $Q_i$.

Then, scalar soft mass squared appear in the scalar potential only through the following combination consistent with the symmetries,
\begin{eqnarray}
V \left( Z + Z^\dag   + Q^\dag_i Q_i\right),
\end{eqnarray}
which give the potential of $Z$ for $Q_i=0$.
The scalar soft mass squared of $Q_i$ is given by
\begin{eqnarray}
\frac{\partial}{\partial Q_i^\dag}\frac{\partial}{\partial Q_i } V\left( Z + Z^\dag  + Q^\dag_i Q_i\right) |_{Q_i=0} = V'\left( Z + Z^\dag\right),
\end{eqnarray}
which vanishes at the vacuum.

We have shown that the scalar soft mass squared is not generated from quantum corrections without yukawa nor gauge coupling constants. Also, as we have discussed, renormalizable yukawa nor gauge interactions alone do not generate soft masses.
Thus, scalar soft mass squared is generated only by corrections involving both yukawa/gauge coupling constants and higher dimensional interactions.
Such corrections are absent at the one-loop level, and are possible only from the two-loop level.
Thus, quantum corrections to scalar soft mass squared are at most as large as
\begin{eqnarray}
\Delta m_{\rm scalar}^2 \sim \frac{\lambda^2}{(16\pi^2)^2} m_{3/2}^2 \left(\frac{\Lambda}{M_*}\right)^n,
\end{eqnarray}
where $\lambda$ is the yukawa or gauge coupling constants.

Finally, we comment on how the above discussion may be invalidated by the regularization of quantum corrections.
We have argued that 
when MSSM soft masses vanish at the tree-level, renormalizable interactions in MSSM do not generate soft masses.
This argument may be invalidated by the regularization.
Consider, for example, the Pauli-Villars regularization.
If Pauli-Villars fields have tree-level soft masses, loop corrections from Pauli-Villars fields generate MSSM soft masses at the one-loop level~\cite{Gaillard:2000fk,Evans:2013uza}.
For the above discussion to be valid, tree-level soft masses of Pauli-Villars fields must be also absent.
In the sequestering based on the brane world, this assumption holds if Pauli-Villars fields also live on the brane of the visible sector.
%In SUSY non-linear sigma models, this assumption holds if Pauli-Villars fields also have Nambu-Goldstone boson natures.

\end{document}